\documentclass[review,12pt]{elsarticle}
\usepackage[english]{babel}
\usepackage{graphicx}

\usepackage{amssymb}
\usepackage{amsthm}
\usepackage{math}
\usepackage{amsfonts}
\usepackage{amsmath}
\usepackage{xcolor}

\usepackage{graphicx}
\usepackage{psfrag}
\usepackage{adjustbox}
\usepackage{multirow}

\usepackage{float}
\usepackage[linesnumbered,ruled,vlined]{algorithm2e}

\journal{Pattern Recognition}

\begin{document}

\begin{frontmatter}

%% Title, authors and addresses

%% use the tnoteref command within \title for footnotes;
%% use the tnotetext command for theassociated footnote;
%% use the fnref command within \author or \address for footnotes;
%% use the fntext command for theassociated footnote;
%% use the corref command within \author for corresponding author footnotes;
%% use the cortext command for theassociated footnote;
%% use the ead command for the email address,
%% and the form \ead[url] for the home page:
%% \title{Title\tnoteref{label1}}
%% \tnotetext[label1]{}
%% \author{Name\corref{cor1}\fnref{label2}}
%% \ead{email address}
%% \ead[url]{home page}
%% \fntext[label2]{}
%% \cortext[cor1]{}
%% \affiliation{organization={},
%%             addressline={},
%%             city={},
%%             postcode={},
%%             state={},
%%             country={}}
%% \fntext[label3]{}

\title{A Network Classification Method based on Density Time Evolution Patterns Extracted from Network Automata}

%% use optional labels to link authors explicitly to addresses:
%% \author[label1,label2]{}
%% \affiliation[label1]{organization={},
%%             addressline={},
%%             city={},
%%             postcode={},
%%             state={},
%%             country={}}
%%
%% \affiliation[label2]{organization={},
%%             addressline={},
%%             city={},
%%             postcode={},
%%             state={},
%%             country={}}

\author[inst1]{Kallil M. C. Zielinski}
\author[ibilce]{Lucas C. Ribas}
\author[inst3]{Jeaneth Machicao}
\author[inst1]{Odemir M. Bruno}

\address[inst1]{S\~{a}o Carlos Institute of Physics,\\ University of S\~{a}o Paulo, S\~{a}o Carlos, SP, Brazil\\}

\address[ibilce]{Institute of Biosciences, Humanities and Exact Sciences, \\ 
São Paulo State University, São José do Rio Preto, SP, Brazil}

\address[inst3]{Computer Engineering Department, \\
Polytechnic School of the University of São Paulo, São Paulo, SP, Brazil}

\begin{abstract}
%% Text of abstract
Network modeling has proven to be an efficient tool for many interdisciplinary areas, including social, biological, transport, and many other real world complex systems. In addition, cellular automata (CA) are a formalism that has been studied in the last decades as a model for exploring patterns in the dynamic spatio-temporal behavior of these systems based on local rules. Some studies explore the use of cellular automata to analyze the dynamic behavior of networks, denominating them as network automata (NA). Recently, NA proved to be efficient for network classification, since it uses a time-evolution pattern (TEP) for the feature extraction. However, the TEPs explored by previous studies are composed of binary values, which does not represent detailed information on the network analyzed. Therefore, in this paper, we propose alternate sources of information to use as descriptor for the classification task, which we denominate as density time-evolution pattern (D-TEP) and state density time-evolution pattern (SD-TEP). We explore the density of alive neighbors of each node, which is a continuous value, and compute feature vectors based on histograms of the TEPs. Our results show a significant improvement compared to previous studies at five synthetic network databases and also seven real world databases. Our proposed method demonstrates not only a good approach for pattern recognition in networks, but also shows great potential for other kinds of data, such as images. 
\end{abstract}

%%Graphical abstract
%\begin{graphicalabstract}
%\includegraphics{grabs}
%\end{graphicalabstract}

%%Research highlights
%\begin{highlights}
%\item Research highlight 1
%\item Research highlight 2
%\end{highlights}

\begin{keyword}
%% keywords here, in the form: keyword \sep keyword
Complex Networks \sep Cellular Automata \sep Network Automata \sep Pattern Recognition
%% PACS codes here, in the form: \PACS code \sep code
%\PACS 0000 \sep 1111
%% MSC codes here, in the form: \MSC code \sep code
%% or \MSC[2008] code \sep code (2000 is the default)
%\MSC 0000 \sep 1111
\end{keyword}

\end{frontmatter}

%% \linenumbers

%% main text
\section{Introduction}

Network science have stood out as a representative model of complex systems due to their multidisciplinary character and ability to represent elements of a system and their interactions. Once the elements of the systems and their connectivity are established, networks can be used to model many real world applications, such as: Natural Phenomena \cite{barjatia2016, abe2006, donges2009},  Biology \cite{sporns2009, rain2001, jeong2000}, Social \cite{dodds2003, palla2005, scabini2021}, Physical \cite{agus2013, carmi2009}, etc. A classical example is the connectivity between routers and computers by cables and optical fibers, forming the well known Internet \cite{Mata2020}.

With the technological advances and the introduction of the \textit{big data} phenomenon, there is an increasing demand for pattern recognition methods that deal with large amount of highly complex data, being difficult to process it by the reductionist approach \cite{miranda2016}. Thus, networks are used to model these data, turning possible to find patterns that are difficult to be identified using other approaches. This is because of the correlation between the elements of the network's topology, in which is feasible to extract many metrics and use these features to distinguish network types \cite{costa2007}. Moreover, network modeling for pattern recognition tasks has recently proven to achieve promising results in many applications, such as computer vision \cite{backes2013, casanova2010, scabini2020, ribas2020} and natural language processing \cite{ipsic2016, das2022, machicao2016}. 

On the other hand, \textit{Cellular Automata} (CA) are a class of discrete models that are also capable of representing complex systems and have turned into an important tool for complexity analysis of spatio-temporal patterns in studies by Wolfram \cite{wolfram1983, Wolfram2002}. These patterns are formed by evolutions in the state of a cell based in the observation of its neighborhood according to a defined set of rules. Despite CAs being originally designed for regular tesselations, like an image, most real world applications are based in irregular tesselations, such as graphs or networks.

The integration between complex networks and cellular automata is already known in the literature, and is denominated \textit{Network Automata} (NA) \cite{smith2011}. Over the network tesselation, the cells of a CA are represented by the network's nodes and the cell neighborhood is represented by the edges of a node. The dynamic behavior of a network automaton can be visualized by a time-evolution pattern (TEP), where the states of each cell are represented at each rule iteration, some approaches such as Miranda et al \cite{miranda2016} have proposed to use this representation, where intrinsic network properties can be extracted. Results from Marr \& Hütt \cite{marr2005, marr2009, hutt2012} shows that there is a strong relation between the degree distribution of the network and the entropy of TEPs. However, these studies are focused on the structural and evolution analysis of networks, therefore, the use of NAs in pattern recognition tasks is an underexplored research field which has great potential. 

Miranda et al \cite{miranda2016} developed a NA approach, denominated \textit{Life-Like Network Automata} (LLNA), as a tool for network analysis for pattern recognition applications. This method uses a family of CAs inspired by the rules of Life-Like, an extension of the Conway's Game of Life \cite{gardner1970}, and it measures the patterns produced by the TEPs to further use it as features for the network classification. While Miranda et al \cite{miranda2016} used entropy measures, Ribas et al \cite{ribas2019} observed the binary patterns of the TEP. In another study, Machicao et al \cite{machicao2016} analysed the efficiency of the method in an authorship attribution task. However, these previous works consider the TEP only by using their binary values representing the state of each node (dead or alive), and could be further improved if taken more information. For this reason, it is important to put more efforts in finding an optimal source to extract information in order be further used as features for the network classification task.

\begin{figure}[!t]
    \centering
    \includegraphics[width=0.8\textwidth]{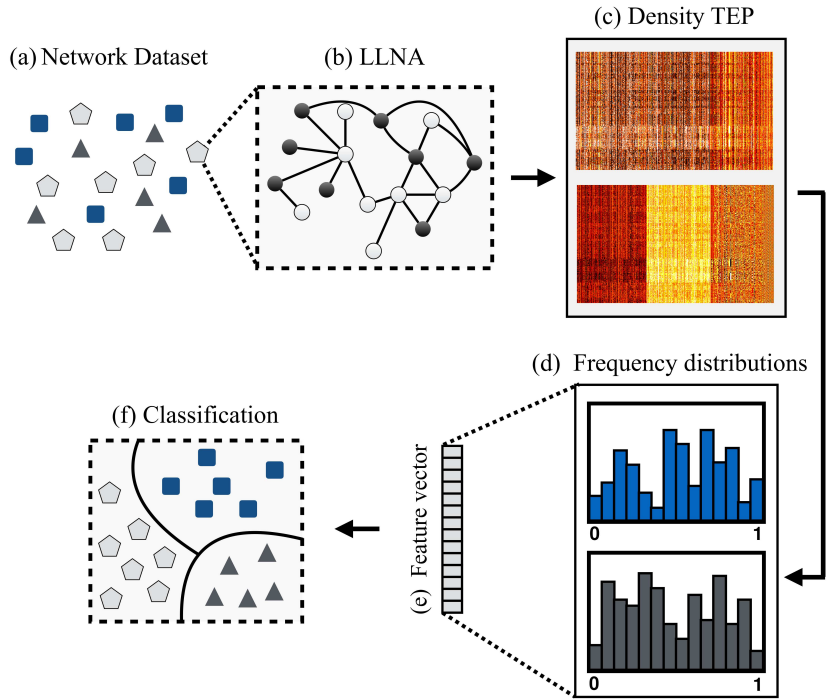} 
    \caption{General scope of a given network classification task. Given a network dataset, each network sample receives a value that determines its initial state, which can be zeros or ones (black and white vertices). From this initial configuration, the D-TEPs and SD-TEPs are extracted by evolving the automata with a determined rule for a fixed number of timesteps and observing the densities of alive neighbors of each cell at each iteration. Histograms are then extracted for different number of bins \binsize and combined into a feature vector to be further used by a classifier algorithm.}
    \label{fig:general_scope}
\end{figure}

In this paper, we propose two improved sources of information in order to extract feature vectors, which we denominate as density time-evolution pattern (D-TEP) and state density time-evolution pattern (SD-TEP). In our approach, instead of using zeros and ones, we use the density of alive neighbors during each iteration, which is a continuous value. The density value was already used in the previous papers \cite{miranda2016, machicao2018, ribas2020} as a mean to calculate the state of each cell, therefore, we explore the use of this value for feature extraction purposes. The general scope of this study can be visualized in Fig \ref{fig:general_scope}.

For the evaluation of the proposed approach, we used five datasets of synthetic networks and also seven datasets of real world networks. The first synthetic dataset is composed of four different models: small-world \cite{Watts1998}, random \cite{erdos1959}, geographical \cite{Waxman1988} and scale-free \cite{barabasi1999}, and the remainder of the synthetic datasets are based in the models from the latter one.
As for the real world applications, they are composed of one social network and six metabolic network datasets. Regarding the social network, it is composed of social interactions between users of two online social networks (Twitter and Google+) \cite{snapnets}. Finally, the six metabolic network databases are obtained from the Kyoto Encyclopedia of Genes and Genomes database (KEGG) \cite{Kanehisa2016}.

%This paper is structured as follows: Section \ref{sec:background} shows the fundamentals of the \textit{Life-Like Network Automata} approach. An overview of methodological procedures of our proposed approach is presented in Section \ref{sec:methodology}. Configurations of experiments are detailed in Section \ref{sec:experiments}. Finally, Sections \ref{sec:results} and \ref{sec:conclusions} shows the results and final considerations, respectively, of this study. 

\section{Background} \label{sec:background}

\subsection{Network Automata}

Network automata are a generalization of CAs, in which the tesselation is a network, where the vertices represent the cells, the edges represent the neighborhood and its behavior is defined by a local rule.

Although the concept of NA was formally defined in 2011 by Smith et al \cite{smith2011}, the first studies that integrated both areas of networks and CAs began in the 1990 by Watts \cite{watts1999}. Later, Tomassini et al \cite{tomassini2005} discuss the use of evolutionary algorithms to evolve small-world CA. Marr \& Hütt \cite{marr2005, marr2009} extracted entropy measures from the TEPs obtained by evolving networks through the use of CAs in order to further classify them according to Wolfram's four classes \cite{wolfram1983}. All of these studies focuses on analyzing changes in the topological behavior of these networks. Miranda et al \cite{miranda2016} in 2016 introduced a NA based on the Life-Like CA family called the Life-Like Network Automata (LLNA), aimed for network classification tasks.

\subsection{Life-Like Network Automata}

Formally, a CA \ca is defined by the tuple \catuple, where: \tesselation is the tesselation, composed by a set of cells $c_i$; \spaceSet is the finite set of states, in this case it is binary (alive or dead), then the notation $s(c_i,t)=1$ defines the state of cell $c_i$ at the timestep $t$; \initState is defined as the initial condition of the CA, where a value is attributed for each cell at $t=0$. Generally, a random value is calculated for each cell, in Miranda et al \cite{miranda2016}, it was considered that the proportion of alive cells in the initial state is approximately 50\% in order to avoid bias and to guarantee randomness; \neighbor is the neighborhood of each cell and, finally, \transition is the transition function, also called the rule of the CA, which defines its dynamic evolution.

In the two-dimensional CA Game of Life \cite{gardner1970}, a future state of a cell $c_i$ is dependent on its own present state and also the state of its neighboring cells. For example, each cell $c_i$ in the two dimensional grid receives a state, being dead ($s(c_i,t)=0$) or alive ($s(c_i,t)=1$). Every cell interacts with its eight neighbours, which are the cells that are directly horizontally, vertically or diagonally adjacent. At each timestep, the following rule is applied:

\begin{itemize}
    \item Any alive cell with fewer than two alive neighbors will die of loneliness;
    \item Any alive cell with more than three alive neighbors will die of overcrowding;
    \item Any alive cell with two or three alive neighbors will survive;
    \item Any dead cell with three alive neighbors will born.
\end{itemize}

This rule is denoted as B3/S23, this means that a cell is born if has 3 neighbors alive and survives if there is 2 or 3 neighbors alive. 
Similarly, Life-Like rules can be characterized by the notation B$x$/S$y$ (e.g. B4/S0, B02/S56, B567/S09, etc), such that $x = \{x_0,x_1,...,x_n|x_i \in \mathbb{N}, 0\leq x_i \leq 8\}$ and $y = \{y_0,y_1,...,y_n|y_i \in \mathbb{N}, 0\leq y_i \leq 8\}$. It is important to note that the combination of all these condition results on a rule space of $2^{(9+9)} = 262,144$ Life-Like rules.

The LLNA proposed by Miranda et al \cite{miranda2016} is inspired in a two dimensional CA that uses the Life-Like rules. In LLNA, the tesselation \tesselation is given by the network topology, where each vertex is represented by a cell $c_i$. The classic Life-Like rules, such as the Game of Life \cite{gardner1970}, are generally defined over a regular tesselation, with a fixed number of neighbors corresponding to a Moore neighborhood $r=9$. Meanwhile, networks are irregular tesselations, i.e. the number of neighbors of each cell is not fixed, and instead is restricted to the connectivity (neighborhood) of each vertex of the network. Therefore, in order to apply Life-Like rules in a network, Miranda et al \cite{miranda2016} defined the LLNA transition function in terms of the density of alive cells in a neighborhood $\rho$. This correspondence is given below:

\begin{equation}
    \rho(c_i,t) = \frac{1}{k_i}\sum_j s(c_i,t)A(i,j) 
    \label{eq:density}
\end{equation} where $A$ is the adjacency matrix of the graph: If there is an edge between node $i$ and node $j$ in the network, then $A(i,j)= 1$, or $0$ otherwise, and $k_i$ is the degree of $c_i$.

Miranda et al \cite{miranda2016} considered the notation of the Life-Like rules $Bx/Sy$ and a neighborhood with size $r=9$, and defined the LLNA transition function \transition: $s(c_i, t) \xrightarrow{} s(c_i,t+1)$ as: 

\begin{equation}
    s(c_i, t+1) = 
    \begin{cases}
    1, & \text{ if } s(c_i,t) = 0 \text{ and } \frac{x_x}{r} \leq \rho(c_i,t) < \frac{x_x+1}{r} \longrightarrow \text{ Born (B) }  \\
    1, & \text{ if } s(c_i,t) = 1 \text{ and } \frac{y_y}{r} \leq \rho(c_i,t) < \frac{y_y+1}{r} \longrightarrow \text{ Survive (S) }  \\
    0, & \text{ otherwise }
    \end{cases}
    \label{eq: llnaConditions}
\end{equation}

For the sake of illustration, consider a LLNA evolving with rule B3/S23, and a cell $c_i$ with the following parameters: $s(c_i,t)=0$ and $\rho(c_i,t) = \frac{1}{3}$. Since $s(c_i,t)=0$, then it activates the born rule and Equation \ref{eq: llnaConditions} will be as: 
$$s(c_i, t+1) = \begin{cases}
1, & \text{ if }  \frac{3}{9} \leq \rho(c_i,t) < \frac{4}{9} \\
0, & \text{ otherwise }
\end{cases}$$
then $s(c_i, t+1) = 1$.

As a second example, consider another cell $c_j$ with the following parameters: $s(c_j,t) = 1$ and $\rho(c_j,t) = \frac{5}{6}$. This time, since $s(c_j,t) = 1$, we calculate $s(c_j,t+1)$ from the survive rule $S23$:
$$s(c_j, t+1) = \begin{cases}
1, & \text{ if }  \frac{2}{9} \leq \rho(c_j,t) < \frac{3}{9} \text{ or } \frac{3}{9} \leq \rho(c_j,t) < \frac{4}{9} \\
0, & \text{ otherwise }
\end{cases}$$
then $s(c_j, t+1) = 0$.

\subsection{Time-evolution pattern}

A time-evolution pattern (TEP) can be described as a concatenation of the states of the cells in an automaton over time. In mathematical terms, it is a two-dimensional matrix \tep of size $(T \times N)$, where $N$ is the number of cells of the automaton and $T$ is the number of timesteps in which the NA is evolved. The TEP is defined as: 

\begin{equation}
    \tep =  
    \begin{bmatrix}
    s(c_1,0) & s(c_2, 0) & ... & s(c_N,0) \\
    s(c_1,1) & s(c_2, 1) & ... & s(c_N,1) \\
    \vdots & \vdots & \vdots \\
    s(c_1, T) & s(c_2,T) & ... & s(c_N,T)
    \end{bmatrix}
    \label{eq:tep}
\end{equation}

Since the states of each cell in the NA are binaries, therefore the extracted TEPs are composed of chains of zeros and ones, representing their cell's states (alive or dead), e.g. 0100101110110101.

\subsection{Feature extraction from TEPs}

As presented in the previous section, a TEP is obtained, after evolving a NA for a fixed number of timesteps $T$. These TEPs containing binary values can be used to calculate properties that are further used as features for classification purposes \cite{miranda2016, ribas2019}. Miranda et al and Machicao et al \cite{miranda2016, machicao2016} used the distribution of histograms of three distinct measures: Shannon entropy \cite{Shannon1948}, the word length and the Lempel-Ziv complexity \cite{estevez2015}. The combination of these measures were used as features. 

Since the LLNA rule space has a considerable number of rules (262,144), Miranda et al \cite{miranda2016} also made a rule selection procedure. This was done by searching for the set of rules that performs better in the pre-classification task. However, a set of selected rules from one dataset may not perform so well in another dataset. Therefore, for each classification task, it was necessary to apply the rule selection procedure using a small part of the data, so that it could be find a set of rules that achieved the best results on this particular dataset. After that, the reported performance of each rule was analyzed in the remaining data using these pre-selected rules. 

Ribas et al \cite{ribas2019} performed a significant improvement of the classification performance by taking the binary patterns of the TEP, and denominated as Life-Like Network Automata - Binary Pattern (LLNA-BP). This method consisted of taking a fixed number $D$ of binary numbers, and with a sliding window of size $D$, it takes the concatenated binary numbers inside each window and transforms into its decimal representation. The distributions of these decimal numbers are then calculated into a histogram of size $2^D$. Ribas et al \cite{ribas2019} proposed two feature extraction methods for the classification tasks: (I) Global histogram, which takes the histogram of all values in the TEP; and (II) Degree histogram, that calculates the histogram separately for columns of the TEP that correspond to cells of a specific degree.

Following the same procedures of the approaches studied \cite{miranda2016, ribas2019}, the NA is initialized with random values (dead or alive) so that the proportion is almost 50\%. One important point is that, once the NA is iterated, their first iterations are usually discarded because they represent the transient period \transient, which is usually a small value (e.g. 20 iterations). This consideration is also in compass with the dynamic theory. 

However, these approaches are based in TEPs that contains the binary state representation of each cell, and despite the good results obtained in these studies, a better representation, with more detailed information, is needed. We observed that the LLNA transition function \transition is based on the density of alive neighbors of each cell, which is defined by Equation \ref{eq:density} and is a continuous value that seems to have more information about what is happening during one cell iteration than its own state value. Therefore, in this paper, we propose two  different approaches to build the TEP, which uses the density value instead of the binary state.

\section{Proposed Time Evolution Patterns} \label{sec:TEPs}

This section details our novel approach to build TEPs, which are named as the density-time evolution pattern (D-TEP) and the state density-time evolution pattern (SD-TEP). 

\subsection{Density-time evolution pattern}

A density time-evolution pattern (D-TEP) is a concatenation of the densities of alive neighbors of the automaton's cells over time. A D-TEP is defined as:

\begin{equation}
    \dtep = 
    \begin{bmatrix}
    \rho(c_1,0) & \rho(c_2, 0) & ... & \rho(c_N,0) \\
    \rho(c_1,1) & \rho(c_2, 1) & ... & \rho(c_N,1) \\
    \vdots & \vdots & \vdots \\
    \rho(c_1, T) & \rho(c_2,T) & ... & \rho(c_N,T)
    \end{bmatrix}
\end{equation}

where $\rho(c_i, t)$ is the density of alive neighbors of a cell $c_i$ at a timestep $t$, which is a continuous value between 0 and 1.

Although the D-TEP is a richer source of information in comparison to the TEP, the LLNA evolution is given by the state transition of its cells, which can be represented by a TEP, and therefore, we propose a combination between both TEP and D-TEP, which is denominated as state density-time evolution pattern (SD-TEP).

\subsection{State density-time evolution pattern}

 We also propose an alternative approach that scales the continuous values of the density, and it interpolates between positive and negative values based on the state of the cells. Thus, for instance, it considers a positive density value if the cell's state is 1, and a negative density if the state is 0. To obtain the SD-TEP, it is important to first obtain the D-TEP (previous section). We denominate this approach as state density-time evolution pattern (SD-TEP), and it is defined as: 

\begin{equation}
    \sdtep = \dtep\odot(2\tep-1)
\end{equation}

where \dtep and \tep are the already calculated D-TEP and TEP, respectively, and $\odot$ is the element-wise product between two matrices with same dimension. Thus, this results in a matrix with weighted values between -1 and 1, depending on the state value of the cells. 

\begin{figure}[!htpb]
\psfrag{NW}{\hspace{-1.8em}\scriptsize{Network}}
\psfrag{RD}{\hspace{-1em}\scriptsize{Random}}
\psfrag{SW}{\hspace{-1.5em}\scriptsize{Small World}}
\psfrag{SF}{\hspace{-1.5em}\scriptsize{Scale Free}}
\psfrag{GG}{\hspace{-1.5em}\scriptsize{Geographic}}
\psfrag{BT}{\hspace{-2em}\scriptsize{TEP}}
\psfrag{DT}{\hspace{-2em}\scriptsize{D-TEP}}
\psfrag{SDT}{\hspace{-2em} \scriptsize{SD-TEP}}
    \centering
    \includegraphics[width=\textwidth]{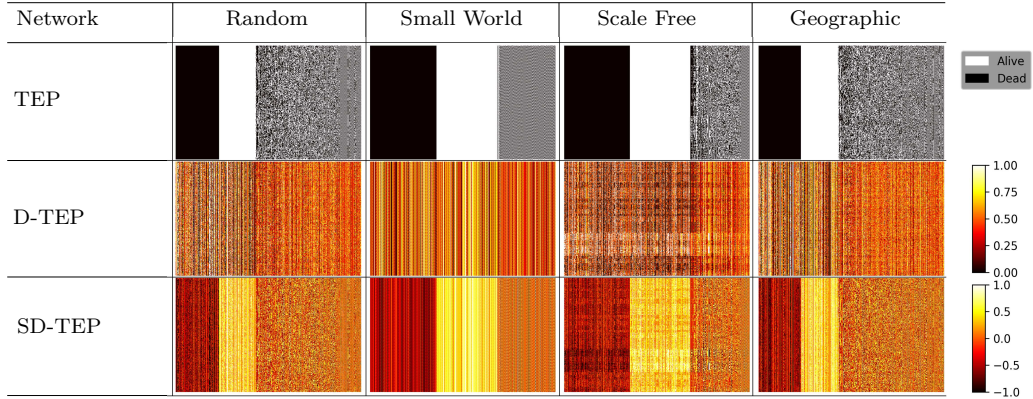}
    \caption{Comparison between the original, density and state density TEPs for each different network class. The parameters used to generate these images were: Number of nodes $N = 500$, average degree $\overline{k} = 4$, number of timesteps $T=350$, and these dynamics were observed for rule B1357-S02468. The columns of each TEP are ordered from left to right by the entropy value.}
    \label{fig:tep-dtep-comparison}
\end{figure}

Figure \ref{fig:tep-dtep-comparison} shows a visual comparison between the TEP, D-TEP and the SD-TEP for random, small-world, scale-free and geographic networks. The TEP is composed of binary values,thus its values are composed of two different colors: black (0) and white (1). In comparison with the TEP, the D-TEP is represented with a collor pallete, with values ranging between 0 and 1. Finally, the SD-TEP can be visualized as the D-TEP weighted by the TEP, and its values range between -1 and 1. The difference between its dynamic behaviors is most noticeable in areas where the TEP show few dynamic transition. For example, the black regions of the TEP are composed of the network's nodes that reached a stable value of 0 over the automaton evolution, while the white regions are composed of a stable value of 1 for the network node. Both regions present a null entropy dynamic, that is, its state is not changing over time during the iterations. The same networks are presented in the D-TEP and the SD-TEP, yet the same nodes that reached stability in the TEP now present more detailed behavior, since the same network shows a wider color palette, ranging from 0 to 1 in the D-TEP and from -1 to 1 in the SD-TEP. As for the latter, we can observe the same low entropy regions, since this TEP carries the information of the states in the network's nodes. This demonstrates a significant difference for representing the automaton.

In summary, both D-TEP and SD-TEP shows richer details mainly in regions where the original TEP present steady pattern, due to the fact that a cell state may not modify its value, but if its neighborhood is changing over time, the density of that cell is also varying. 

\subsection{Feature Extraction of D-TEPs} \label{sec:dtep-measure}

In this paper, we consider three different measures for the feature extraction procedure: A global value histogram, a degree histogram and also a temporal histogram. All of these measures were tested with different number of bins \binsize. 
That is, a dictionary with \binsize intervals is initialized, and for all nodes at every iteration in the D-TEPs, we count the occurrences of values that range in a given interval.

\subsubsection{Global Histogram}

This is similar to the global binary histogram measure in Ribas et al \cite{ribas2019}, the difference is that instead of count the binary occurrences, we simply count the number of occurrences of different density intervals varying from 0 to 1. The mathematical notation of this property is given in Equations \ref{eq:dtepgh} and \ref{eq:densityfunction}.

\begin{equation}
    \dtepgh(\binindex) = \sum_{i=1}^N \sum_{t=0}^T f(\dtep(i,t), \binindex),  
\\
    \label{eq:dtepgh}
\end{equation}
where 
\begin{equation}
    f(\dtep(i,t), \binindex) =  \begin{cases}
    1  & \text{if }  \frac{\binindex}{\binsize} \leq \dtep(i,t) \leq \frac{\binindex+1}{\binsize} \\ 
    0 & \text{otherwise}
    \label{eq:densityfunction}
    \end{cases}
\end{equation}
and $\binindex \in [0,\binsize-1]$.

Also, we can define it as a feature vector notation: 

\begin{equation}
    \dtepmeangh = \begin{bmatrix}
    \dtepgh(0) \\
    \dtepgh(2) \\
    \vdots \\
    \dtepgh(L-1)\\
    \end{bmatrix} 
    \label{eq:dtepmeangh}
\end{equation}

This histogram has information of all the occurrences of values of a given density interval, but since the sum of all occurrences varies with the size of the network, we normalize this histogram in order to turn it invariant to the network size.

\subsubsection{Degree Histogram}

A complex network has many nodes with different degrees $k$, which is the number of neighbors of that specific node. This property makes possible to split histogram calculation for nodes with same degree $k$, which gives us information about patterns observed for similar nodes of the network. In this way, we can define a degree histogram as:

\begin{equation}
    \dtepdh(\binindex, \degree{}) = \sum_{i=1}^N \sum_{t=0}^T f(\dtep(i,t), \binindex) \cdot \delta(k,k_i)
\end{equation}

where \begin{equation}
\delta(x,y) = 
\begin{cases}
1 & \text{if } x = y \\
0 & \text{otherwise }
\end{cases}
\end{equation}, $f(\dtep(i,t), b)$ is defined in Equation \ref{eq:densityfunction}, $\binindex \in [0, \binsize-1]$ and $k_i$ is the degree of node $i$. Note that the delta function in the equation ensures that only nodes with degree $k$ will be calculated on the histogram. Also, similarly to the global histogram, the degree histogram is also normalized to ensure invariance to the network size.

Once a degree histogram \dtepdh is calculated for each degree $k$ of the network, then we propose to represent a single histogram as the average of all histograms for the network characterization: 

\begin{equation}
    \dtepmeandh = \mu \begin{pmatrix}
    \begin{bmatrix}
    \dtepdh(\binindex,\degree{1}) \\
    \dtepdh(\binindex,\degree{2}) \\
    \vdots \\
    \dtepdh(\binindex,\degree{max})\\
    \end{bmatrix}
    \end{pmatrix}
    \label{eq:dtepmeandh}
\end{equation}

where $k_{max}$ is the maximum degree of the network.

\subsubsection{Temporal Histogram}

A single time evolution pattern contains information of the network nodes at each iteration. Similarly to the degree histogram, we can also split the histogram calculation for each time step of the D-TEP. This will give us information about patterns observed for each evolution step of the automaton. A temporal histogram can be defined as:

\begin{equation}
    \dtepth(\binindex, t) = \sum_{i=1}^N f(\dtep(i,t), \binindex)
\end{equation}

where $f(\dtep(i,t), b)$ is defined in Equation \ref{eq:densityfunction}, $\binindex \in [0, \binsize-1]$, and $t \in [0, T]$ with $T$ being the number of time steps the automaton has evolved. This histogram is also normalized for network size invariance.
And similarly to the degree histogram of the previous section, we propose a single histogram for the temporal property as the average of all histograms at each time step.

\begin{equation}
    \dtepmeanth= \mu \begin{pmatrix}\begin{bmatrix}
    \dtepth(\binindex,1) \\
    \dtepth(\binindex,2) \\
    \vdots \\
    \dtepth(\binindex,T)\\
    \end{bmatrix} \end{pmatrix}
    \label{eq:dtepmeanth}
\end{equation}

It is also possible to combine the global, degree and density histogram in order to obtain a better representation of the network. This can be done by concatenating the three feature vectors obtained:

\begin{equation}
    \dtepcat = [\dtepmeangh, \dtepmeandh, \dtepmeanth]
    \label{eq:dtepcat}
\end{equation}

\subsubsection{Feature Extraction of SD-TEPs}

The measures suggested for the feature extraction procedure of the D-TEP in Section \ref{sec:dtep-measure} can also be applied for the SD-TEP. Therefore, we can generalize equations \ref{eq:dtepmeangh}, \ref{eq:dtepmeandh} and \ref{eq:dtepmeanth} into the SD-TEP, obtaining a global histogram \sdtepmeangh, a degree histogram \sdtepmeandh and a temporal histogram \sdtepmeanth. Also, similar to Equation \ref{eq:dtepcat}, we can concatenate these features: 

\begin{equation}
    \sdtepcat = [\sdtepmeangh, \sdtepmeandh, \sdtepmeanth]
\end{equation}

\subsection{Proposed Feature Vector}

Our suggested approach consists of combining histograms with different size of bins in order to obtain an optimal discriminatory representation. We illustrate the difference of histogram sizes in Figure \ref{fig:histogram}, which shows two different networks and their global histograms of sizes 20 and 40, where for each bin sizes, there are different feature vectors to represent the network. So in order to find an optimal representation for the classification task, we prepared an experiment to explore this parameter by combining histograms with different number of bins, as shown below:

\begin{figure}
\psfrag{SW}{\hspace{-2.5em}\scriptsize{Small-world network}}
\psfrag{SF}{\hspace{-2.5em}\scriptsize{Scale-free network}}
\psfrag{W16}{\hspace{-2.5em}\scriptsize{Small-world 20 bins}}
\psfrag{W32}{\hspace{-2.5em}\scriptsize{Small-world 40 bins}}
\psfrag{B16}{\hspace{-2.3em}\scriptsize{Scale-free 20 bins}}
\psfrag{B32}{\hspace{-2.3em}\scriptsize{Scale-free 40 bins}}
    \centering
    \includegraphics[width=\textwidth]{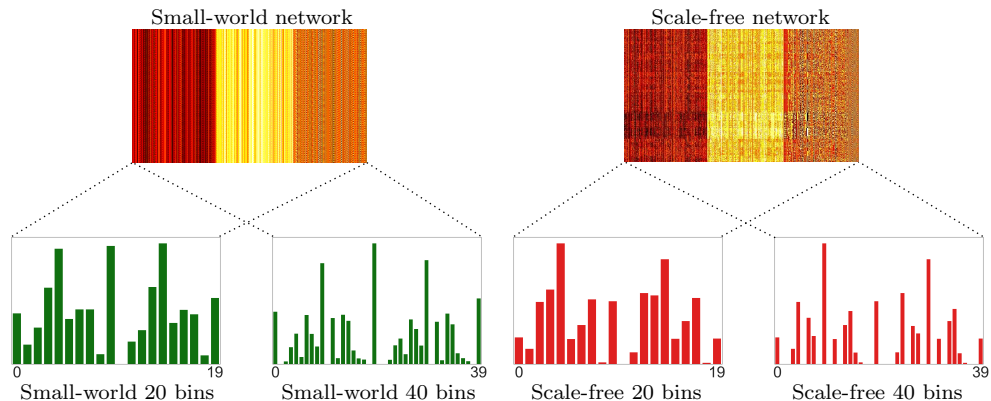}
    \caption{Global histogram representation of small-world and scale-free networks for SD-TEP representation for different bin sizes.}
    \label{fig:histogram}
\end{figure}

\begin{equation}
    \begin{aligned}
    \vec{\Theta}^{\dtep}_{[\binsize_1,\binsize_2,...,\binsize_n]} = [\vec{\Upsilon}^\dtep_{\binsize_1}, \vec{\Upsilon}^\dtep_{\binsize_2}, ... , \vec{\Upsilon}^\dtep_{\binsize_n}] & \xrightarrow{} & \text{ D-TEP } \\
    \vec{\Theta}^{\sdtep}_{[\binsize_1,\binsize_2,...,\binsize_n]} = [\vec{\Upsilon}^\sdtep_{\binsize_1}, \vec{\Upsilon}^\sdtep_{\binsize_2}, ... , \vec{\Upsilon}^\sdtep_{\binsize_n}] & \xrightarrow{} & \text{ SD-TEP } \\
    \end{aligned}
\end{equation}

Finally, we also propose a third signature consisting on concatenating both D-TEP and SD-TEP feature vectors of the three proposed histograms.

\begin{equation}
    \dtepsdtepcomb_{[\binsize_1, \binsize_2, ..., \binsize_n]} = \big[\vec{\Theta}^{\dtep}_{\binsize_1, \binsize_2, ..., \binsize_n} , \vec{\Theta}^{\sdtep}_{\binsize_1, \binsize_2, ..., \binsize_n} \big]
    \label{eq:combinedVector}
\end{equation}

The proposed feature vector combines the information of histogram of different sizes, and, consequently, different information about the D-TEP and SD-TEP behavior.

\section{Experimental Configuration}\label{sec:experiments}

\subsection{Databases} \label{sec:databases}

To conduct the experiments, we used five synthetic network databases and also seven real world network databases. These same databases were used in Miranda et al \cite{miranda2016} and Ribas et al \cite{ribas2020}, so that we could compare the results.

\begin{itemize}
    \item \textit{4-models synthetic-database}: This database has synthetic networks generated according to 4 distinct network classes: Random \cite{erdos1959}, small-world \cite{Watts1998}, scale-free \cite{barabasi1999} and geographical \cite{Waxman1988}. Each class has networks generated with distinct number of nodes ($N=\{500,1000,1500,2000\})$ and also mean degree ($\langle k \rangle = \{4,6,8,10,12,14,16,18\}$), with a total of 28 combinations between the elements of $N$ and $\langle k \rangle$. For each combination it was generated 100 networks, resulting in a total of 2800 networks for each model, so the database has 11200 networks for all of the 4 classes;
    \item \textit{Scalefree-synthetic-database}: This database is composed of scale-free synthetic models generating according to the models proposed by Barabási \& Albert \cite{barabasi1999} and from Dorogovstev \& Mendes \cite{Dorogovtsev2003}. The Barabasi models consisted of four different classes according to the power law parameter $\alpha = \{0.5,1.0,1.5,2.0\}$. Therefore, the \textit{scalefree-synthetic-database} has five classes (four from Barabási and one from Mendes) and each class contains 100 networks with $N=1000$ and $\langle k \rangle = 8$;
    \item \textit{Noisy-synthetic-database}: Using the models from the \textit{4-models synthetic database} and the \textit{scalefree-synthetic-database}, this database consisted of modifying the topology of the networks by the addition or removal of edges. We adopted three different values of percentage of modified edges of the network: $\sigma = \{10\%, 20\%,30\%\}$. As the value of $\sigma$ increases, more the topology of the network changes. A value of $\sigma = 10\%$ means that $5\%$ of edges will be added to the network while $5\%$ will be removed. This database has 8 classes, with 100 networks for each class; 
    \item \textit{Social-networks-database}: There are networks from the SNAP (\textit{Stanford Network Analysis Project}) platform \cite{snapnets}. There are two classes: Google+ and Twitter, and every class contains 50 networks. 
    \item \textit{Metabolic-networks-database}: Composed of metabolic networks constructed with the substrate-product network model \cite{Zhao2006}. This database is based on several biochemical reactions of organisms, which were obtained from the Kyoto Encyclopedia of Genes and Genomes database \cite{Kanehisa2016} (KEGG). The networks were generated according to a model that considered metabolites as vertices and its chemical reactions as the edges. We subdivided these networks into seven different classification schemes, as detailed below:
        \begin{itemize}
            \item \textit{kingdom-database}: Contains species from the \textit{eukaryota} domain, represented by the four kingdoms: \textit{animal, plant, fungi} and \textit{protist}. Each one of the four classes contains 40 networks, resulting in 160 network samples;
            \item \textit{Animal-database}: Represented by 14 samples of four different classes: \textit{mammal, bird, fish} and \textit{insect};
            \item \textit{Fungi-database}: This database has 15 networks for each of its four classes: \textit{saccharomycetes, sordariomycetes, eurotiomycetes} and \textit{basidiomycetes};
            \item \textit{Plant-database}: Contains 27 networks split into 3 different classes: \textit{Monocots, Green Algae} and \textit{Eudicots}. Each class has 9 network samples;
            %\item \textit{Protist-database}: Contains four classes: \textit{Amoebozoa, Alveolates, Stramenopiles} and  \textit{Euglenozoa}, each class has 5 samples;
            \item \textit{Firmicutes-Bacilis-database}: There are four different classes: \textit{Bacillus, Staphilococcus, Streptococcus} and \textit{Lactobacillus} with 122, 76, 133 and 83 samples, respectively, which means that this database is not balanced;
            \item \textit{Actinobacteria-database}: Also an unbalanced database, with three classes: \textit{Mycobacterium, Corynebacterium} and \textit{Streptomyces}, with 60, 86 and 53 samples, respectively;
        \end{itemize}
    %\item \textit{Metabolic-rule-selection-database}: In order to choose the best LLNA for each classification problem, we generated another seven databases containing the metabolic networks described. Therefore, \textit{kingdom-selection-database} has 9 samples for each class, the \textit{animal-selection-database}, \textit{fungi-selection-database}, \textit{protist-selection-database} and \textit{plant-selection-database} contains 2 samples of each class, and both the \textit{firmicutes-Bacillis-selection-database} and \textit{actinobacteria-selection-database} comprises 10 networks samples of each class.
\end{itemize}

\subsection{LLNA configuration parameters}

The purpose of our proposed method is to improve the classification accuracy obtained from the previous works. Therefore, it is important to use the same parameters used in Miranda et al \cite{miranda2016} and Ribas et al \cite{ribas2019}, in order to avoid divergences and compare all the approaches fairly. The LLNA was evolved for a number of $T=350$ timesteps, with the initial configuration \initState setup randomly using a normal distribution of 50\% of cells alive. Also, the rules that were selected for evaluation of the databases in the previous works were the same used in this study.

According to the first LLNA proposal research \cite{miranda2016}, rule B135678-S03456 was used for both \textit{4-models synthetic-database} and \textit{noisy-synthetic-database}. The \textit{4-models + $\langle k \rangle$ synthetic-database} used rule B01678-S0457 while rule B0157-S457 was considered by the \textit{scale-free synthetic-database}. The \textit{social-database} used rule B0167-S246. Now, as a result of the rule-selection procedure made in Ribas et al \cite{ribas2019}, rules B02345678-S123468, B023468-S01468, B04-S1468, B0468-S0467, B0236-S123567, B0468-S0458, B1237-S267 were used for the \textit{Kingdom, Animal, Fungi, Plant, Protist, Firmicutes-Bacillis} and \textit{Actibacteria-databases} respectively.

\subsection{Classification Experiments}

In order to follow the same classification setup presented in the previous works, we used a 10-fold cross-validation strategy, dividing every database into 10 equal subsets, where one is used for validation and the 9 other subsets for training. This algorithm tests all combinations of train-test data split and return the mean value of all the accuracies obtained. The split method is not deterministic, so we apply this cross-validation procedure for 100 times. The classification algorithm used was SVM (Support Vector Machines), which uses decision boundaries and hyperplanes in order to find an optimal hyperplane that guarantees maximal separation between two classes \cite{hearst1998}.

\section{Results}\label{sec:results}

Here we report the results obtained in the classification task for the databases presented in Section \ref{sec:databases}. First, we analyzed the efficiency of the method using different number of histogram bins \binsize, while also studying the performance of combining features generated by different bin sizes. Once chosen the best signature, we compared the classification performance of our proposed approach with the previous studies \cite{miranda2016, ribas2019}.

\subsection{Feature vector selection}

In order to choose the best features to apply in the classification procedure, we analyzed the performance of eight different histogram size values: $\binsize = \{20,40,60,80,100, 120,140,160\}$, and also the combination of two of the selected values. It is important to detail that we did not choose to combine more than two values of \binsize in order to avoid having a large number of features for the classification task, which can impact the efficiency of our method.

We chose the social dataset to be the baseline for comparison of the metrics discussed in Section \ref{sec:dtep-measure}: Global histogram for the D-TEP \dtepmeangh and SD-TEP \sdtepmeangh, degree histograms \dtepmeandh and \sdtepmeandh and the temporal histograms \dtepmeanth and \sdtepmeanth. The metric that has the higher classification accuracy was the combined feature vector \dtepsdtepcomb defined in Equation \ref{eq:combinedVector} (all of the results from every measure are shown in Table S-1 of the supplementary materials). Figure \ref{fig:barplot_social} shows the accuracies obtained for feature vector $\dtepsdtepcomb_\binsize$ for histogram bin combinations of the values $\binsize = \{20,40,60,80,100\}$ in order to simplify the visualization. The combinations that show the best performing histogram sizes are vectors $\dtepsdtepcomb_{\{40,100\}}$, achieving $92.5 \pm 0.50 \%$, and vector $\dtepsdtepcomb_{\{60,100\}}$, achieving $92.5 \pm 0.92\%$. Apart from the social database the feature vector $\dtepsdtepcomb_\binsize$ were also tested in the databases presented in Section \ref{sec:databases}, and the complete results for synthetic and real databases are presented in Tables S-2 and S-3 of the supplementary materials, respectively.

\begin{figure}
    \psfrag{acc}{\hspace{-1.5em}\scriptsize{Accuracy}}
    \psfrag{bset}{\hspace{-3em}\scriptsize{Combination of bin sizes for \dtepsdtepcomb}}
    \centering
    \includegraphics[width=\textwidth]{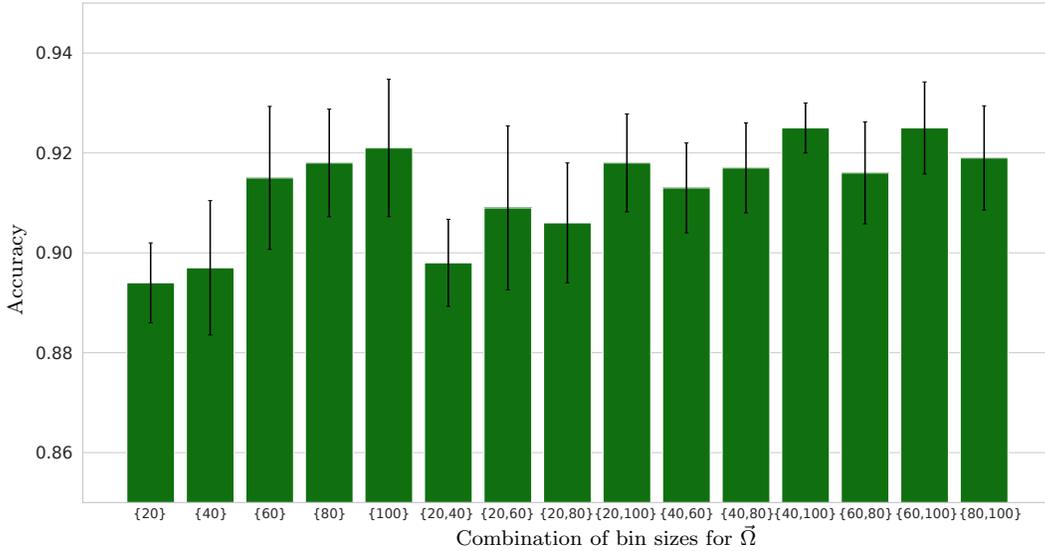}
    \caption{Accuracies for social database using different number of bins}
    \label{fig:barplot_social}
\end{figure}

In general, synthetic databases are easier to classify, since they are computationally generated. The results show that the method performs very well on synthetic databases independent of the combination values of \binsize, the combinations that performed best in the social database, which is $\binsize = \{40,100\}$ and $\{60,100\}$ also performed very well in the synthetic databases. Finally Table \ref{tab:accuracies_real} shows the accuracies and standard deviations of both these feature vectors in the real databases, in which is possible to note that generally they perform well on these databases.

\begin{table}[!htpb]
\centering
\caption{Accuracies and standard deviations (in percentage) of different combinations of histogram sizes for the vectors $\vec{\Omega}_{\{40,100\}}$ and $\vec{\Omega}_{\{60,100\}}$ in the real world databases.}
\resizebox{\textwidth}{!}{
\begin{tabular}{lcccccccccccc}
\hline
\multirow{2}{1em}{$\{\binsize_1,\binsize_2\}$} & \multicolumn{2}{c}{Kingdom}     & \multicolumn{2}{c}{Animal}                        & \multicolumn{2}{c}{Fungi}                         & \multicolumn{2}{c}{Plant}                         & \multicolumn{2}{c}{Firmicutes-Baccilis}           & \multicolumn{2}{c}{Actinobacteria}                \\
                               & Acc   & \multicolumn{1}{c}{Std} & \multicolumn{1}{c}{Acc} & \multicolumn{1}{c}{Std} & \multicolumn{1}{c}{Acc} & \multicolumn{1}{c}{Std} & \multicolumn{1}{c}{Acc} & \multicolumn{1}{c}{Std} & \multicolumn{1}{c}{Acc} & \multicolumn{1}{c}{Std} & \multicolumn{1}{c}{Acc} & \multicolumn{1}{c}{Std} \\ \hline
\{40,  100\}          & 96.24 & 0.33 & 99.71  & 0.57 & 80.43 & 4.66 & 81.33 & 6.00 & 96.06      & 0.35     & 97.68   & 0.28   \\
\{60,  100\}          & 96.24 & 0.35 & 100.00 & 0.00 & 81.00 & 4.38 & 79.58 & 2.12 & 95.73      & 0.34     & 97.65   & 0.29   \\
\end{tabular}}
\label{tab:accuracies_real}
\end{table}

\subsection{Comparison with previous methods}

In this section, we detail the efficiency of our method by directly comparing the two best feature vectors $\dtepsdtepcomb_{\{40,100\}}$ and $\dtepsdtepcomb_{\{60,100\}}$ with the previous methods that were used in the literature: \textit{Life-like network automata - Binary patterns} (LLNA-BP) \cite{ribas2019}, the original LLNA \cite{miranda2016} and also classical network measurements. LLNA-BP use the global and degree histograms of binary patterns obtained from the TEP as features. LLNA use Shannon entropy, word length and Lempel-ziv complexity of the TEP. Finally, the network measures used in classification were: mean degree, degree distributions, correlations, distances, path lengths, hierarchical and spectral measures, transitivity, clustering coefficient, among others. The accuracies from these methods were obtained from Ribas et al \cite{ribas2019}. 

Table \ref{tab:accuracies_comparison} presents the results of each approach evaluated in the real and synthetic network databases detailed in Section \ref{sec:databases}. For the synthetic databases, our both feature vectors have the same results. They perform better than LLNA in most cases, and it also performs better than LLNA-BP in the \textit{scale-free and noise 20\%} databases. As for the real-world databases, our method significantly outperformed the other methods in the \textit{animal} (99.71 $\pm$ 0.57 for $\dtepsdtepcomb_{\{40,100\}}$ and 100.0 $\pm 0.00$ for $\dtepsdtepcomb_{\{60,100\}}$), \textit{fungi} (80.43 $\pm$ 4.66 for $\dtepsdtepcomb_{\{40,100\}}$ and 81.00 $\pm$ 4.38 for $\dtepsdtepcomb_{\{60,100\}}$) and \textit{plant} (81.33 $\pm$ 6.00 for $\dtepsdtepcomb_{\{40,100\}}$ and 79.58 $\pm$ 2.12 for $\dtepsdtepcomb_{\{60,100\}}$) databases. Note that most of our results not only improve the accuracy, but also the standard deviation, since our method reduces the variation in the classification task. In most cases, our approach is better in terms of standard deviation, even if the accuracy does not outperform the other methods.
In general, our results show that our method improved at the classification task in many of the studied databases when compared to the previous methods.

\begin{table}[!htpb]
\caption{Comparison of our proposal approach accuracy and standard deviation (\%) with the \textit{Life-Like network automata - Binary patters} (LLNA-BP) \cite{ribas2019} and the original LLNA approach \cite{miranda2016}.}
\centering
\resizebox{\textwidth}{!}{
\begin{tabular}{llccccc}
\hline
\multicolumn{2}{l}{Databases}                    & \multicolumn{5}{c}{Methods}                                                              \\
Type                       & Name                & $\vec{\Omega}_{\{60,100\}}$ (Ours) & $\vec{\Omega}_{\{40,100\}}$ (Ours) & LLNA-BP            & LLNA              & Structural measures \\ \hline
\multirow{5}{6em}{Synthetic} & 4-models           & 100.0 $\pm$ 0.00 & 100.0 $\pm$ 0.00          & 100.0 $\pm$ 0.00   & 99.99 $\pm$ 0.00  & 100.0.0 $\pm$ 0.00  \\
                           & Scale-free      & 100.0 $\pm$ 0.00     & 100.0 $\pm$ 0.00          & 99.52 $\pm$ 0.19   & 98.30 $\pm$ 0.07  & 96.20 $\pm$ 0.20    \\
                           & Noise 10\%      & 100.0 $\pm$ 0.00     & 100.0 $\pm$ 0.00          & 100.0.0 $\pm$ 0.00 & 99.98 $\pm$ 0.00  & 100.0.0 $\pm$ 0.00  \\
                           & Noise 20\%       & 100.0 $\pm$ 0.00    & 100.0 $\pm$ 0.00          & 99.98 $\pm$ 0.00   & 99.97 $\pm$ 0.01  & 100.0.0 $\pm$ 0.00  \\
                           & Noise 30\%        & 99.75 $\pm$ 0.00   & 99.75 $\pm$ 0.00          & 99.99 $\pm$ 0.00   & 99.95 $\pm$ 0.01  & 100.0.0 $\pm$ 0.00  \\
                           \hline
\multirow{7}{6em}{Real}      & Social          & 92.5 $\pm$ 0.50    & 92.5 $\pm$ 0.92           & 93.40 $\pm$ 0.92   & 92.00 $\pm$ 1.00  & 88.00 $\pm$ 2.00    \\
                           & Kingdom        & 96.24 $\pm$ 0.35     & 96.24 $\pm$ 0.33          & 97.44 $\pm$ 3.98   & 93.10 $\pm$ 5.38  & 96.61 $\pm$ 4.33    \\
                           & Animal        & 100.0 $\pm$ 0.00      & 99.71 $\pm$ 0.57          & 84.87 $\pm$ 15.25  & 77.25 $\pm$ 16.29 & 83.71 $\pm$ 15.29   \\
                           & Fungi         & 81.00 $\pm$ 4.38      & 80.43 $\pm$ 4.66          & 76.17 $\pm$ 17.45  & 54.58 $\pm$ 19.38 & 54.90 $\pm$ 15.39   \\
                           & Plant         & 79.58 $\pm$ 2.12      & 81.33 $\pm$ 6.00          & 74.81 $\pm$  5.64  & 69.70 $\pm$ 4.67  & 54.19 $\pm$ 9.17    \\
                           & Firmicutes-Baccilis & 95.73 $\pm$ 0.34  & 96.06 $\pm$ 0.35          & 98.30 $\pm$ 1.17   & 84.63 $\pm$ 2.00  & 95.67 $\pm$ 0.59    \\
                           & Actinobacteria  & 97.65 $\pm$ 0.29    & 97.68 $\pm$ 0.28          & 95.13 $\pm$ 1.22   & 91.48 $\pm$ 1.60  & 93.16 $\pm$ 0.70    \\ \hline
\end{tabular}}
\label{tab:accuracies_comparison}
\end{table}

\section{Final Considerations}\label{sec:conclusions}

This paper presents an improved feature extraction method based on \textit{Life-like network automata}. We denominate these properties as density time-evolution pattern (D-TEP) and state density time-evolution pattern (SD-TEP), which are later used for feature extraction in the context of pattern recognition. These properties uses the density of alive neighbors of each node and how they evolve in time. In comparison to the original TEP, which uses sequences of zeros and ones to represent the state evolution of the network's nodes, D-TEP uses decimal numbers between 0 and 1 and SD-TEP between -1 and 1, depending on the state of the cell. These values contain a much detailed representation of the network automata dynamic behavior. 

By using global, degree and temporal histograms of these values, we demonstrate the robustness of our method in five synthetic databases and also seven real-world databases. First, we evaluated the optimal size of bins and combinations of the histograms, in order to generally perform well in all of the databases. Then, experimental results using the proposed signature confirmed the efficiency of out method while comparing to previous methods that makes use of the LLNA and also structural network measures. Our method demonstrates a significantly higher performance mainly in the real databases that the previous approaches performed poorly. This demonstrates that even in challenging scenarios, the dynamic behavior of the proposed TEPs gives a signature that leads to an effective interpretation of the network that is being analysed. 

Although the results show a significant upgrade in comparison to previous methods, we emphasize that our method focuses on investigate if using the density of alive cells for feature extraction would improve the LLNA performance. For this reason, we use the same rules that were classified as best discriminators in the previous studies, and our following steps are to make a rule selection procedure using D-TEPs or SD-TEPs, which can probably improve the results. Therefore, we present a sophisticated method that presents a better discriminator nature of networks.

\section*{Acknowledgements}
K. M. C. Zielinski acknowledge support from CAPES (grants \#88887. 631085/ 2021-00). L. C. Ribas acknowledge support from São Paulo Research Foundation (FAPESP) (grants \#2021/07289-2). J. Machicao. is grateful for the support from FAPESP (Grant \#2020/03514-9). O. M. Bruno acknowledges support from CNPq (Grant \#307897/2018-4) and FAPESP (grants \#2018/22214-6 and \#2021/08325-2). The authors are also grateful to the NVIDIA GPU Grant Program.

%% The Appendices part is started with the command \appendix;
%% appendix sections are then done as normal sections
\appendix

 \bibliographystyle{elsarticle-num} 
 \bibliography{cas-refs}

%% else use the following coding to input the bibitems directly in the
%% TeX file.

% \begin{thebibliography}{00}

% %% \bibitem{label}
% %% Text of bibliographic item

% \bibitem{}

% \end{thebibliography}
\end{document}

\begin{frontmatter}

%% Title, authors and addresses

%% use the tnoteref command within \title for footnotes;
%% use the tnotetext command for theassociated footnote;
%% use the fnref command within \author or \address for footnotes;
%% use the fntext command for theassociated footnote;
%% use the corref command within \author for corresponding author footnotes;
%% use the cortext command for theassociated footnote;
%% use the ead command for the email address,
%% and the form \ead[url] for the home page:
%% \title{Title\tnoteref{label1}}
%% \tnotetext[label1]{}
%% \author{Name\corref{cor1}\fnref{label2}}
%% \ead{email address}
%% \ead[url]{home page}
%% \fntext[label2]{}
%% \cortext[cor1]{}
%% \affiliation{organization={},
%%             addressline={},
%%             city={},
%%             postcode={},
%%             state={},
%%             country={}}
%% \fntext[label3]{}

\title{A Network Classification Method based on Density Time Evolution Patterns Extracted from Network Automata - Supplementary Materials}

%% use optional labels to link authors explicitly to addresses:
%% \author[label1,label2]{}
%% \affiliation[label1]{organization={},
%%             addressline={},
%%             city={},
%%             postcode={},
%%             state={},
%%             country={}}
%%
%% \affiliation[label2]{organization={},
%%             addressline={},
%%             city={},
%%             postcode={},
%%             state={},
%%             country={}}

\author[inst1]{Kallil M. C. Zielinski}

\author[inst1]{Lucas C. Ribas}
\author[inst3]{Jeaneth Machicao}
\author[inst1]{Odemir M. Bruno}

\affiliation[inst1]{organization={Scientific Computing Group},%Department and Organization
            addressline={Institute of Physics of Sao Carlos - University of Sao Paulo}, 
            city={Sao Carlos}, 
            state={SP},
            country={Brazil}}

%%Graphical abstract
%\begin{graphicalabstract}
%\includegraphics{grabs}
%\end{graphicalabstract}

%%Research highlights
%\begin{highlights}
%\item Research highlight 1
%\item Research highlight 2
%\end{highlights}

\end{frontmatter}

\section{Auxiliary Tables}

\begin{table}[]
\centering
\caption{Accuracies and standard deviations (in percentage) of different histogram measures for different number of bins and the combination of all histograms for the social database.}
\resizebox{\textwidth}{!}{
\begin{tabular}{lcccccccccccccc}
\hline
\multirow{}{}{$\{\binsize_1,\binsize_2\}$} & \multicolumn{2}{c}{\dtepmeangh}                    & \multicolumn{2}{c}{\sdtepmeangh}                 & \multicolumn{2}{c}{\dtepmeandh}              & \multicolumn{2}{c}{\sdtepmeandh}           & \multicolumn{2}{c}{\dtepmeanth}         & \multicolumn{2}{c}{\sdtepmeanth}      & \multicolumn{2}{c}{$\vec{\Omega}_\binsize$}                    \\
                                   & \multicolumn{1}{c}{Acc} & \multicolumn{1}{c}{Std} & \multicolumn{1}{c}{Acc} & \multicolumn{1}{c}{Std} & \multicolumn{1}{c}{Acc} & \multicolumn{1}{c}{Std} & \multicolumn{1}{c}{Acc} & \multicolumn{1}{c}{Std} & \multicolumn{1}{c}{Acc} & \multicolumn{1}{c}{Std} & \multicolumn{1}{c}{Acc} & \multicolumn{1}{c}{Std} & \multicolumn{1}{c}{Acc} & \multicolumn{1}{c}{Std} \\ \hline
$\{20\}$                           & 88.40                   & 0.66                    & 88.70                   & 0.64                    & 88.00                   & 0.77                    & 89.10                   & 0.54                    & 87.90                   & 1.04                    & 88.40                   & 1.02                    & 89.40                   & 0.80                    \\
$\{40\}$                           & 89.00                   & 1.34                    & 91.10                   & 0.70                    & 89.40                   & 0.66                    & 90.10                   & 1.45                    & 88.80                   & 1.33                    & 90.50                   & 1.57                    & 89.70                   & 1.35                    \\
$\{60\}$                           & 90.70                   & 1.55                    & 91.20                   & 0.75                    & 90.70                   & 1.55                    & 90.70                   & 1.73                    & 90.90                   & 0.83                    & 91.10                   & 0.83                    & 91.50                   & 1.43                    \\
$\{80\}$                           & 91.50                   & 0.81                    & 91.60                   & 0.66                    & 91.40                   & 0.92                    & 91.60                   & 0.66                    & 90.70                   & 1.85                    & 91.20                   & 1.25                    & 91.80                   & 1.08                    \\
$\{100\}$                          & 90.80                   & 1.08                    & 91.70                   & 1.10                    & 90.80                   & 0.98                    & 92.30                   & 1.00                    & 90.70                   & 0.78                    & 91.60                   & 0.92                    & 92.10                   & 1.37                    \\
$\{120\}$                          & 91.50                   & 0.67                    & 92.10                   & 0.64                    & 91.60                   & 0.66                    & 92.10                   & 1.00                    & 91.20                   & 0.87                    & 92.30                   & 1.19                    & 92.30                   & 0.78                    \\
$\{140\}$                          & 90.60                   & 0.80                    & 92.20                   & 1.02                    & 90.70                   & 1.19                    & 92.10                   & 0.46                    & 90.00                   & 1.48                    & 92.40                   & 1.11                    & 92.10                   & 0.54                    \\
$\{160\}$                          & 90.00                   & 0.45                    & 91.90                   & 1.45                    & 89.70                   & 0.78                    & 91.90                   & 0.77                    & 89.90                   & 0.54                    & 92.40                   & 0.64                    & 91.60                   & 0.66                    \\
$\{20 ,40\}$                       & 89.00                   & 1.34                    & 88.40                   & 1.36                    & 89.90                   & 1.04                    & 89.10                   & 1.14                    & 90.80                   & 0.87                    & 90.20                   & 1.47                    & 89.80                   & 0.87                    \\
$\{20,  60\}$                      & 91.60                   & 0.49                    & 89.60                   & 0.80                    & 91.60                   & 0.66                    & 89.80                   & 0.87                    & 91.40                   & 0.80                    & 91.20                   & 1.08                    & 90.90                   & 1.64                    \\
$\{20,  80\}$                      & 91.50                   & 0.67                    & 88.90                   & 1.14                    & 91.50                   & 0.92                    & 90.20                   & 1.17                    & 91.00                   & 1.48                    & 91.00                   & 0.89                    & 90.60                   & 1.20                    \\
$\{20,  100\}$                     & 90.70                   & 0.64                    & 89.60                   & 1.56                    & 91.70                   & 0.90                    & 91.60                   & 1.02                    & 91.00                   & 0.77                    & 90.70                   & 2.33                    & 91.80                   & 0.98                    \\
$\{20,  120\}$                     & 91.10                   & 0.83                    & 91.40                   & 1.50                    & 91.30                   & 1.27                    & 92.30                   & 0.46                    & 90.50                   & 1.12                    & 92.10                   & 1.04                    & 91.90                   & 1.22                    \\
$\{20,  140\}$                     & 90.30                   & 1.00                    & 90.90                   & 1.92                    & 91.10                   & 1.22                    & 92.50                   & 0.92                    & 90.40                   & 1.11                    & 92.10                   & 0.78                    & 92.40                   & 0.92                    \\
$\{20,  160\}$                     & 90.10                   & 0.70                    & 91.40                   & 1.20                    & 91.00                   & 0.63                    & 92.30                   & 1.00                    & 90.40                   & 1.36                    & 92.20                   & 1.02                    & 92.40                   & 0.80                    \\
$\{40,  60\}$                      & 90.10                   & 0.94                    & 89.00                   & 1.67                    & 90.80                   & 1.08                    & 89.50                   & 1.02                    & 91.60                   & 0.80                    & 90.30                   & 1.27                    & 91.30                   & 0.90                    \\
$\{40,  80\}$                      & 90.70                   & 1.85                    & 88.50                   & 1.20                    & 91.10                   & 1.14                    & 91.20                   & 1.08                    & 91.60                   & 0.66                    & 91.80                   & 0.75                    & 91.70                   & 0.90                    \\
\textbf{\{40,  100\} }                    & \textbf{90.10}                   & \textbf{1.04}                    & \textbf{89.60}                   & \textbf{0.80}                    & \textbf{92.20 }                  & \textbf{1.17}                    & \textbf{90.70}                   & \textbf{1.00}                    & \textbf{91.10}                   & \textbf{0.94}                    & \textbf{91.80}                   & \textbf{0.87}                    & \textbf{92.50}                   & \textbf{0.92 }                   \\
$\{40,  120\}$                     & 90.90                   & 0.83                    & 90.60                   & 1.02                    & 90.80                   & 1.08                    & 92.30                   & 0.64                    & 91.60                   & 0.66                    & 91.80                   & 1.40                    & 92.30                   & 1.42                    \\
$\{40,  140\}$                     & 90.80                   & 0.60                    & 90.20                   & 0.87                    & 91.20                   & 0.75                    & 92.20                   & 0.87                    & 90.50                   & 0.92                    & 91.30                   & 1.27                    & 92.20                   & 0.98                    \\
$\{40,  160\}$                     & 90.50                   & 0.67                    & 90.30                   & 1.00                    & 90.70                   & 0.64                    & 91.90                   & 1.14                    & 90.60                   & 0.66                    & 92.10                   & 0.70                    & 92.00                   & 1.00                    \\
$\{60,  80\}$                      & 90.20                   & 1.25                    & 91.10                   & 1.51                    & 91.60                   & 0.92                    & 90.10                   & 1.30                    & 91.30                   & 0.90                    & 91.30                   & 1.42                    & 91.60                   & 1.02                    \\
\textbf{\{60,  100\} }                    & \textbf{90.90 }                  & \textbf{1.14}                    & \textbf{90.40}                   & \textbf{1.85}                    & \textbf{91.50}                   & \textbf{0.67}                    & \textbf{90.90}                   & \textbf{1.14}                    & \textbf{91.20}                   & \textbf{0.98}                    & \textbf{91.40}                   & \textbf{1.28}                    & \textbf{92.50}                   & \textbf{0.50}                    \\
$\{60,  120\}$                     & 90.60                   & 0.66                    & 89.80                   & 1.25                    & 91.30                   & 0.64                    & 91.80                   & 1.40                    & 91.40                   & 0.80                    & 92.30                   & 0.64                    & 92.40                   & 0.92                    \\
$\{60,  140\}$                     & 90.50                   & 0.67                    & 90.30                   & 1.00                    & 90.80                   & 0.87                    & 92.30                   & 1.27                    & 90.30                   & 0.78                    & 91.30                   & 1.27                    & 92.20                   & 1.08                    \\
$\{60,  160\}$                     & 90.70                   & 0.46                    & 90.30                   & 1.00                    & 91.00                   & 0.00                    & 92.00                   & 1.34                    & 90.60                   & 0.49                    & 92.30                   & 1.73                    & 91.80                   & 1.72                    \\
$\{80,  100\}$                     & 90.80                   & 1.72                    & 92.00                   & 1.26                    & 90.90                   & 1.04                    & 90.90                   & 1.76                    & 91.30                   & 0.78                    & 91.50                   & 1.36                    & 91.90                   & 1.04                    \\
$\{80,  120\}$                     & 90.70                   & 0.78                    & 90.10                   & 1.76                    & 91.80                   & 0.75                    & 91.30                   & 1.68                    & 91.40                   & 0.66                    & 91.70                   & 1.10                    & 92.10                   & 1.04                    \\
$\{80,  140\}$                     & 91.10                   & 0.54                    & 91.20                   & 1.25                    & 91.10                   & 0.70                    & 92.50                   & 1.02                    & 90.70                   & 0.78                    & 92.50                   & 0.81                    & 92.40                   & 0.66                    \\
$\{80,  160\}$                     & 90.60                   & 0.49                    & 90.40                   & 1.20                    & 90.80                   & 0.40                    & 92.10                   & 1.14                    & 90.80                   & 0.40                    & 92.50                   & 1.91                    & 92.00                   & 1.00                    \\
$\{100,  120\}$                    & 91.00                   & 1.34                    & 91.90                   & 0.94                    & 90.90                   & 0.54                    & 91.50                   & 1.75                    & 90.90                   & 0.54                    & 92.30                   & 1.55                    & 92.30                   & 1.00                    \\
$\{100 , 140\}$                    & 90.90                   & 1.14                    & 91.20                   & 1.17                    & 90.90                   & 0.83                    & 92.20                   & 1.17                    & 90.60                   & 0.49                    & 92.40                   & 0.92                    & 91.90                   & 1.97                    \\
$\{100,  160\}$                    & 90.80                   & 0.40                    & 90.80                   & 1.40                    & 90.90                   & 0.30                    & 91.50                   & 1.69                    & 91.00                   & 0.45                    & 92.20                   & 0.80                    & 92.00                   & 1.48                    \\
$\{120,  140\}$                    & 90.70                   & 1.27                    & 91.60                   & 0.92                    & 90.80                   & 0.60                    & 91.60                   & 1.11                    & 91.00                   & 0.45                    & 92.40                   & 1.11                    & 92.10                   & 1.04                    \\
$\{120,  160\}$                    & 90.70                   & 0.90                    & 91.80                   & 0.60                    & 91.00                   & 0.45                    & 91.30                   & 1.00                    & 90.70                   & 0.46                    & 91.60                   & 1.28                    & 91.90                   & 0.70                    \\
$\{140,  160\}$                    & 90.90                   & 0.83                    & 92.00                   & 0.63                    & 90.80                   & 0.40                    & 91.70                   & 1.10                    & 90.80                   & 0.40                    & 91.90                   & 0.77                    & 91.70                   & 0.90                    \\ \hline
\end{tabular}
}
\label{tab:accuracies_social}
\end{table}

\begin{table}[]
\centering
\caption{Accuracies and standard deviations (in percentage) of different combinations of histogram sizes for the combined vector $\vec{\Omega}_\binsize$ in the synthetic databases.} 
\resizebox{0.78\textwidth}{!}{
\begin{tabular}{lrrrrrrrrrr}
\hline
\multirow{}{}{$\{\binsize_1,\binsize_2\}$} & \multicolumn{2}{c}{\multirow{}{}{4-Models}}     & \multicolumn{2}{c}{\multirow{}{}{Scale-free}}   & \multicolumn{6}{c}{Noise}                                                                                                                                 \\
                               & \multicolumn{2}{c}{}                              & \multicolumn{2}{c}{}                              & \multicolumn{2}{c}{10\%}                          & \multicolumn{2}{c}{20\%}                          & \multicolumn{2}{c}{30\%}                          \\ \cline{2-11} 
                               & \multicolumn{1}{c}{Acc} & \multicolumn{1}{c}{Std} & \multicolumn{1}{c}{Acc} & \multicolumn{1}{c}{Std} & \multicolumn{1}{c}{Acc} & \multicolumn{1}{c}{Std} & \multicolumn{1}{c}{Acc} & \multicolumn{1}{c}{Std} & \multicolumn{1}{c}{Acc} & \multicolumn{1}{c}{Std} \\ \hline
$\{20\}$                       & 99.87                   & 0.00                    & 100.00                  & 0.00                    & 100.00                  & 0.00                    & 100.00                  & 0.00                    & 100.00                  & 0.00                    \\
$\{40\}$                       & 100.00                  & 0.00                    & 100.00                  & 0.00                    & 100.00                  & 0.00                    & 100.00                  & 0.00                    & 100.00                  & 0.00                    \\
$\{60\}$                       & 100.00                  & 0.00                    & 100.00                  & 0.00                    & 100.00                  & 0.00                    & 100.00                  & 0.00                    & 99.87                   & 0.00                    \\
$\{80\}$                       & 100.00                  & 0.00                    & 100.00                  & 0.00                    & 99.87                   & 0.00                    & 100.00                  & 0.00                    & 99.84                   & 0.08                    \\
$\{100\}$                      & 99.61                   & 0.10                    & 100.00                  & 0.00                    & 100.00                  & 0.00                    & 100.00                  & 0.00                    & 99.75                   & 0.00                    \\
$\{120\}$                      & 100.00                  & 0.00                    & 100.00                  & 0.00                    & 99.89                   & 0.07                    & 99.55                   & 0.13                    & 99.86                   & 0.04                    \\
$\{140\}$                      & 100.00                  & 0.00                    & 100.00                  & 0.00                    & 99.99                   & 0.04                    & 99.42                   & 0.11                    & 99.83                   & 0.10                    \\
$\{160\}$                      & 99.98                   & 0.08                    & 100.00                  & 0.00                    & 99.99                   & 0.04                    & 99.55                   & 0.06                    & 99.74                   & 0.04                    \\
$\{20 ,40\}$                   & 99.98                   & 0.08                    & 100.00                  & 0.00                    & 100.00                  & 0.00                    & 100.00                  & 0.00                    & 100.00                  & 0.00                    \\
$\{20,  60\}$                  & 99.98                   & 0.08                    & 100.00                  & 0.00                    & 100.00                  & 0.00                    & 100.00                  & 0.00                    & 99.89                   & 0.04                    \\
$\{20,  80\}$                  & 100.00                  & 0.00                    & 100.00                  & 0.00                    & 99.93                   & 0.06                    & 100.00                  & 0.00                    & 99.87                   & 0.00                    \\
$\{20,  100\}$                 & 100.00                  & 0.00                    & 100.00                  & 0.00                    & 100.00                  & 0.00                    & 100.00                  & 0.00                    & 99.75                   & 0.00                    \\
$\{20,  120\}$                 & 100.00                  & 0.00                    & 100.00                  & 0.00                    & 99.98                   & 0.05                    & 99.66                   & 0.14                    & 99.79                   & 0.11                    \\
$\{20,  140\}$                 & 100.00                  & 0.00                    & 100.00                  & 0.00                    & 100.00                  & 0.00                    & 99.60                   & 0.08                    & 99.87                   & 0.00                    \\
$\{20,  160\}$                 & 100.00                  & 0.00                    & 100.00                  & 0.00                    & 99.98                   & 0.05                    & 99.77                   & 0.11                    & 99.75                   & 0.00                    \\
$\{40,  60\}$                  & 100.00                  & 0.00                    & 100.00                  & 0.00                    & 100.00                  & 0.00                    & 100.00                  & 0.00                    & 99.98                   & 0.05                    \\
$\{40,  80\}$                  & 100.00                  & 0.00                    & 100.00                  & 0.00                    & 99.99                   & 0.04                    & 100.00                  & 0.00                    & 99.86                   & 0.04                    \\
\textbf{\{40,  100\}}          & \textbf{100.00}         & \textbf{0.00}           & \textbf{100.00}         & \textbf{0.00}           & \textbf{100.00}         & \textbf{0.00}           & \textbf{100.00}         & \textbf{0.00}           & \textbf{99.75}          & \textbf{0.00}           \\
$\{40,  120\}$                 & 100.00                  & 0.00                    & 99.95                   & 0.10                    & 100.00                  & 0.00                    & 99.66                   & 0.11                    & 99.86                   & 0.04                    \\
$\{40,  140\}$                 & 99.73                   & 0.05                    & 100.00                  & 0.00                    & 100.00                  & 0.00                    & 99.66                   & 0.11                    & 99.86                   & 0.04                    \\
$\{40,  160\}$                 & 100.00                  & 0.00                    & 100.00                  & 0.00                    & 99.99                   & 0.04                    & 99.77                   & 0.09                    & 99.73                   & 0.08                    \\
$\{60,  80\}$                  & 100.00                  & 0.00                    & 100.00                  & 0.00                    & 100.00                  & 0.00                    & 100.00                  & 0.00                    & 99.86                   & 0.04                    \\
\textbf{\{60,  100\}}          & \textbf{100.00}         & \textbf{0.00}           & \textbf{100.00}         & \textbf{0.00}           & \textbf{100.00}         & \textbf{0.00}           & \textbf{100.00}         & \textbf{0.00}           & \textbf{99.75}          & \textbf{0.00}           \\
$\{60,  120\}$                 & 100.00                  & 0.00                    & 100.00                  & 0.00                    & 100.00                  & 0.00                    & 99.81                   & 0.06                    & 99.81                   & 0.13                    \\
$\{60,  140\}$                 & 100.00                  & 0.00                    & 100.00                  & 0.00                    & 100.00                  & 0.00                    & 99.70                   & 0.10                    & 99.85                   & 0.07                    \\
$\{60,  160\}$                 & 100.00                  & 0.00                    & 100.00                  & 0.00                    & 99.98                   & 0.05                    & 99.75                   & 0.15                    & 99.70                   & 0.10                    \\
$\{80,  100\}$                 & 100.00                  & 0.00                    & 100.00                  & 0.00                    & 100.00                  & 0.00                    & 100.00                  & 0.00                    & 99.75                   & 0.00                    \\
$\{80,  120\}$                 & 99.59                   & 0.08                    & 99.98                   & 0.08                    & 99.98                   & 0.05                    & 99.77                   & 0.13                    & 99.77                   & 0.13                    \\
$\{80,  140\}$                 & 99.48                   & 0.09                    & 100.00                  & 0.00                    & 100.00                  & 0.00                    & 99.70                   & 0.06                    & 99.83                   & 0.10                    \\
$\{80,  160\}$                 & 99.86                   & 0.04                    & 100.00                  & 0.00                    & 100.00                  & 0.00                    & 99.70                   & 0.10                    & 99.71                   & 0.08                    \\
$\{100,  120\}$                & 100.00                  & 0.00                    & 99.98                   & 0.08                    & 99.84                   & 0.08                    & 99.83                   & 0.10                    & 99.81                   & 0.10                    \\
$\{100 , 140\}$                & 100.00                  & 0.00                    & 100.00                  & 0.00                    & 100.00                  & 0.00                    & 99.74                   & 0.07                    & 99.81                   & 0.10                    \\
$\{100,  160\}$                & 100.00                  & 0.00                    & 100.00                  & 0.00                    & 99.99                   & 0.04                    & 99.79                   & 0.08                    & 99.75                   & 0.00                    \\
$\{120,  140\}$                & 100.00                  & 0.00                    & 99.95                   & 0.10                    & 100.00                  & 0.00                    & 99.34                   & 0.19                    & 99.81                   & 0.12                    \\
$\{120,  160\}$                & 99.90                   & 0.12                    & 99.98                   & 0.08                    & 99.94                   & 0.08                    & 99.63                   & 0.00                    & 99.73                   & 0.05                    \\ 
$\{140,  160\}$                & 99.93                   & 0.16                    & 100.00                  & 0.00                    & 100.00                  & 0.00                    & 99.54                   & 0.13                    & 99.73                   & 0.08    \\ \hline               
\end{tabular}
}
\label{tab:accuracies_synthetic}
\end{table}

% Please add the following required packages to your document preamble:
% \usepackage{multirow}
\begin{table}[htpb]
\centering
\caption{Accuracies and standard deviations (in percentage) of different combinations of histogram sizes for the combined vector $\vec{\Omega}_\binsize$ in the real world databases.}
\resizebox{0.95\textwidth}{!}{
\begin{tabular}{lcccccccccccc}
\hline
\multirow{}{}{$\{\binsize_1,\binsize_2\}$} & \multicolumn{2}{c}{Kingdom}    & \multicolumn{2}{c}{Animal}      & \multicolumn{2}{c}{Fungi}      & \multicolumn{2}{c}{Plant}      & \multicolumn{2}{c}{Firmicutes-Baccilis} & \multicolumn{2}{c}{Actinobacteria} \\
                               & Acc            & Std           & Acc             & Std           & \multicolumn{2}{c}{Acc}        & \multicolumn{2}{c}{Acc}        & \multicolumn{2}{c}{Acc}                 & Acc              & Std             \\ \hline
$\{20\}$                       & 98.87          & 0.20          & 97.14           & 0.61          & 59.79          & 1.86          & 63.58          & 8.72          & 95.27               & 0.30              & 97.60            & 0.29            \\
$\{40\}$                       & 98.42          & 0.15          & 99.50           & 1.07          & 67.60          & 6.92          & 67.50          & 5.55          & 94.83               & 0.30              & 97.25            & 0.20            \\
$\{60\}$                       & 96.84          & 0.37          & 99.67           & 0.67          & 77.69          & 2.18          & 67.25          & 6.00          & 95.54               & 0.24              & 97.81            & 0.02            \\
$\{80\}$                       & 96.01          & 0.22          & 99.10           & 0.98          & 71.95          & 2.46          & 69.75          & 5.73          & 95.19               & 0.32              & 97.42            & 0.37            \\
$\{100\}$                      & 95.82          & 0.20          & 99.21           & 0.79          & 78.93          & 3.36          & 81.33          & 2.67          & 95.89               & 0.29              & 97.47            & 0.18            \\
$\{120\}$                      & 95.92          & 0.03          & 99.83           & 0.50          & 79.07          & 4.95          & 77.50          & 5.99          & 94.80               & 0.50              & 97.94            & 0.20            \\
$\{140\}$                      & 96.53          & 0.31          & 98.52           & 0.50          & 83.81          & 2.74          & 73.58          & 5.12          & 94.57               & 0.34              & 97.52            & 0.20            \\
$\{160\}$                      & 95.97          & 0.18          & 98.60           & 0.48          & 78.24          & 5.08          & 73.83          & 8.72          & 94.43               & 0.14              & 97.38            & 0.39            \\
$\{20 ,40\}$                   & 98.32          & 0.33          & 98.31           & 1.12          & 63.74          & 4.39          & 66.92          & 6.17          & 95.35               & 0.22              & 97.38            & 0.19            \\
$\{20,  60\}$                  & 97.45          & 0.02          & 99.33           & 0.82          & 74.79          & 1.92          & 69.83          & 6.08          & 95.21               & 0.37              & 97.77            & 0.14            \\
$\{20,  80\}$                  & 96.07          & 0.34          & 99.83           & 0.50          & 72.10          & 4.39          & 70.17          & 3.85          & 95.89               & 0.29              & 97.69            & 0.27            \\
$\{20,  100\}$                 & 96.14          & 0.35          & 100.00          & 0.00          & 79.10          & 2.87          & 81.25          & 4.66          & 95.93               & 0.29              & 97.69            & 0.20            \\
$\{20,  120\}$                 & 95.77          & 0.21          & 100.00          & 0.00          & 75.17          & 3.79          & 73.75          & 6.16          & 95.11               & 0.40              & 97.73            & 0.27            \\
$\{20,  140\}$                 & 96.90          & 0.16          & 98.64           & 0.46          & 79.43          & 3.85          & 72.75          & 7.02          & 94.51               & 0.30              & 97.33            & 0.31            \\
$\{20,  160\}$                 & 95.76          & 0.40          & 98.90           & 0.72          & 77.74          & 2.81          & 72.83          & 3.17          & 94.59               & 0.46              & 97.11            & 0.45            \\
$\{40,  60\}$                  & 96.93          & 0.39          & 100.00          & 0.00          & 77.05          & 3.71          & 66.75          & 6.44          & 95.68               & 0.23              & 97.78            & 0.14            \\
$\{40,  80\}$                  & 96.07          & 0.25          & 99.36           & 0.79          & 72.05          & 2.89          & 71.17          & 4.22          & 95.52               & 0.38              & 97.77            & 0.13            \\
\textbf{\{40,  100\}}          & \textbf{96.24} & \textbf{0.33} & \textbf{99.71}  & \textbf{0.57} & \textbf{80.43} & \textbf{4.66} & \textbf{81.33} & \textbf{6.00} & \textbf{96.06}      & \textbf{0.35}     & \textbf{97.68}   & \textbf{0.28}   \\
$\{40,  120\}$                 & 95.54          & 0.33          & 100.00          & 0.00          & 77.48          & 5.18          & 75.00          & 4.20          & 95.00               & 0.25              & 97.60            & 0.30            \\
$\{40,  140\}$                 & 96.44          & 0.02          & 98.43           & 0.12          & 82.48          & 3.43          & 71.42          & 6.42          & 94.84               & 0.34              & 97.07            & 0.34            \\
$\{40,  160\}$                 & 95.97          & 0.15          & 98.95           & 0.94          & 78.81          & 3.56          & 72.92          & 4.19          & 94.89               & 0.47              & 96.95            & 0.34            \\
$\{60,  80\}$                  & 95.96          & 0.17          & 99.86           & 0.43          & 72.74          & 4.43          & 68.83          & 4.19          & 95.57               & 0.37              & 97.69            & 0.28            \\
\textbf{\{60,  100\}}          & \textbf{96.24} & \textbf{0.35} & \textbf{100.00} & \textbf{0.00} & \textbf{81.00} & \textbf{4.38} & \textbf{79.58} & \textbf{2.12} & \textbf{95.73}      & \textbf{0.34}     & \textbf{97.65}   & \textbf{0.29}   \\
$\{60,  120\}$                 & 96.05          & 0.34          & 100.00          & 0.00          & 77.93          & 3.78          & 73.92          & 5.12          & 95.02               & 0.39              & 97.77            & 0.13            \\
$\{60,  140\}$                 & 96.57          & 0.48          & 99.52           & 0.73          & 79.76          & 2.83          & 75.08          & 6.95          & 94.81               & 0.35              & 97.52            & 0.34            \\
$\{60,  160\}$                 & 95.77          & 0.31          & 99.40           & 0.73          & 80.14          & 2.06          & 73.42          & 4.91          & 94.59               & 0.48              & 97.43            & 0.41            \\
$\{80,  100\}$                 & 96.31          & 0.21          & 99.55           & 0.69          & 79.90          & 3.56          & 78.75          & 3.40          & 95.90               & 0.28              & 97.55            & 0.30            \\
$\{80,  120\}$                 & 95.46          & 0.35          & 99.52           & 0.73          & 78.43          & 3.37          & 78.17          & 2.95          & 94.95               & 0.31              & 97.60            & 0.22            \\
$\{80,  140\}$                 & 96.22          & 0.25          & 99.12           & 0.72          & 79.21          & 3.12          & 78.58          & 5.36          & 95.22               & 0.35              & 97.16            & 0.22            \\
$\{80,  160\}$                 & 95.73          & 0.24          & 98.60           & 0.80          & 80.81          & 2.78          & 73.25          & 5.93          & 94.92               & 0.27              & 97.03            & 0.27            \\
$\{100,  120\}$                & 95.58          & 0.35          & 100.00          & 0.00          & 77.83          & 3.67          & 76.33          & 3.40          & 94.73               & 0.52              & 97.68            & 0.20            \\
$\{100 , 140\}$                & 96.40          & 0.16          & 98.79           & 0.62          & 85.14          & 5.17          & 78.67          & 4.96          & 94.94               & 0.23              & 97.33            & 0.24            \\
$\{100,  160\}$                & 95.76          & 0.40          & 98.88           & 0.74          & 79.83          & 3.99          & 78.08          & 4.69          & 94.78               & 0.24              & 97.34            & 0.23            \\
$\{120,  140\}$                & 96.95          & 0.02          & 98.43           & 0.12          & 81.90          & 2.95          & 76.67          & 6.45          & 95.05               & 0.29              & 97.16            & 0.50            \\
$\{120,  160\}$                & 95.92          & 0.41          & 98.45           & 0.12          & 79.48          & 3.62          & 73.50          & 6.49          & 94.51               & 0.38              & 97.03            & 0.47            \\
$\{140,  160\}$                & 95.98          & 0.15          & 98.45           & 0.12          & 80.74          & 3.30          & 74.42          & 5.68          & 94.57               & 0.38              & 96.94            & 0.40            \\ \hline
\end{tabular}
}
\label{tab:accuracies_real}
\end{table}